# Tunable quantum critical point and detached superconductivity in Al-doped CrAs

Sungmin Park[a,1], Soohyeon Shin[a,1], Sung-Il Kim[a], Joe D. Thompson[b], and Tuson Park[a,2]

[a]Department of Physics & Center for Quantum materials and Superconductivity Sungkyunkwan University, Suwon 16419, South Korea; [b]Los Alamos National Laboratory, Los Alamos, NM 87545, USA

[1]S. Park and S. Shin contributed equally to this work

[2]To whom correspondence should be addressed: Prof. Tuson Park, [a]Department of Physics & Center for Quantum materials and Superconductivity Sungkyunkwan University, Suwon 16419, South Korea

Tel. +82-31-299-4543, Email: tp8701@skku.edu

Classification:

PHSICAL SCIENCES: Physics

## Significance

Observation of a quantum critical point (QCP) in correlated unconventional superconductors has been key to understanding of the intricate relationship between superconductivity and quantum criticality, which has been often hampered by the dome of superconducting (SC) state that veils the $T$=0 K quantum phase transition. This work demonstrates observation of a QCP in the Al-doped CrAs, where the QCP became completely detached from the dome of superconducting phase. The tuned QCP and its separation from the SC state imply that Cooper pair formation is not mediated solely by critical magnetic fluctuations in CrAs. This work illustrates the potential of using multiple non-thermal parameters to reveal the relationship between superconductivity and a hidden quantum critical point in classes of unconventional superconductors.

## Abstract

The origin of unconventional superconductivity and its relationship to a $T$=0 K continuous quantum phase transition (a quantum critical point, QCP), which is hidden inside the dome of a superconducting state, have long been an outstanding puzzle in correlated superconductors. The observation and tuning of the hidden QCP, which is critical in resolving the mystery, however, has been rarely reported due to lack of ideal systems. The helical antiferromagnet CrAs provides an example in which a dome of superconductivity appears at a pressure where its magnetic transition goes to zero temperature. Here we report the tuning of a projected critical point in CrAs via Al chemical doping (Al-CrAs) and separation of the magnetic critical point from the pressure-induced superconducting phase. When CrAs is doped with Al, its AFM ordering temperature $T_N$ increases from 260 K to 270 K. With applied pressure, $T_N$ decreases and extrapolates to zero Kelvin near 4.5 kbar, which is shifted from 8 kbar for undoped CrAs. A funnel of anomalously enhanced electron scattering and a non-Fermi liquid resistivity underscore an AFM QCP near 4.5 kbar in Al-CrAs. Pressure-induced superconductivity, in contrast, is almost independent of Al doping and forms a dome with essentially the identical maximum $T_c$ and same optimal pressure as in pure CrAs. The clear separation between the tuned AFM QCP and $T_c$ maximum in Al-CrAs suggests that superconductivity is independent of the AFM QCP, illustrating subtleties in the interplay between superconductivity and quantum criticality in correlated electron systems.

Unconventional superconductivity commonly emerges in proximity to a magnetically ordered phase, raising the possibility that critical spin fluctuations may mediate the formation of superconducting (SC) Cooper pairs [1-10]. In the vast majority of these cases, as sketched in Fig. 1(a), the zero-temperature limit of a continuous magnetic phase boundary (a quantum critical point, QCP), is veiled by a dome of superconductivity, making it difficult to prove the interplay between unconventional superconductivity and fluctuations arising from the presumed QCP. Nevertheless, unusual normal state properties of these materials above their dome of superconductivity imply that a magnetic QCP may remain a viable concept even below $T_c$ in various classes of unconventional superconductors, such as those based on Fe and Cu as well as heavy fermion compounds, in which there is an intricate interplay among intertwined order parameters [11-14]. More definitive evidence for the connection between quantum criticality and unconventional superconductivity may come from experiments showing that a superconducting phase is pinned to or detached from tunable QCPs [15]. When a QCP is moved, as sketched in Fig 1(b), the superconducting phase will be pinned to the tuned QCP if Cooper pairing is produced by the critical quantum fluctuations, while it can detach from the QCP if the two phenomena are independent of each other, therefore providing a stringent test to resolve the potential relationship between QCP and unconventional superconducting state in strongly correlated systems.

CrAs, with MnP orthorhombic structure, orders in a non-collinear helimagnetic structure below $T_N$=260 K, which is accompanied by a discontinuous lattice expansion along the crystalline *b*-axis and contraction along *a*- and *c*-axes [16 -18]. The crystal structure remains unchanged through the transition but its cell volume, dominated by expansion along *b*, is larger below 260 K.

With initial applied pressure, the coupled magnetic and structural transitions move to lower temperatures. Though weak diamagnetism and zero resistance appear already at pressures near 3 kbar, the highest $T_c$ occurs near the critical pressure of 8 kbar where the coupled magnetic/structural transition is projected to zero Kelvin and electrical resistivity deviates from the Landau-Fermi liquid $T^2$ dependence, indicating a helical AFM QCP hidden below the dome of pressure-induced superconductivity in CrAs [19-22]. Recent neutron scattering suggest that the non-Fermi liquid behavior arises from a nearly second-order helical magnetic phase transition that is accompanied by a first-order isostructural transition [22]. Even though these experiments provide circumstantial evidence for a close relationship between quantum criticality and superconductivity, the complexity of simultaneous magnetic and isostructural transitions and the possibility of electronic phase separation in polycrystalline CrAs [23] cloud a straightforward connection between criticality and superconductivity.

Here we show that a projected critical point in CrAs is successfully shifted by Al chemical substitution and the pressure-induced superconducting phase is detached from the tuned magnetic critical point. Slight Al-doping increases $T_N$ from 260 K to 270 K but pressure rapidly suppresses $T_N$ of Al-CrAs to zero Kelvin near 4.5 kbar ($=P_C$), giving a suppression rate that is nearly two times faster than that of pure CrAs. The residual resistivity as well as the temperature coefficient of resistivity peak near $P_C$ shows that the projected critical point is shifted from 8 kbar for pure CrAs to 4.5 kbar by Al doping. Contrary to the tunable critical point, the maximum $T_c$ of pressure-induced SC state remains near 8 kbar, showing that the SC dome is detached from the shifted QCP. These discoveries evidence that superconductivity in CrAs is produced in spite of the QCP, not because of it. The unambiguous demonstration of detached superconductivity from the QCP in Al-CrAs illustrates that tuning via non-thermal control parameters can provide an alternative route to probe the intricate relationship between a hidden critical point and surrounding superconductivity.

**Results and discussion**

Figure 2(a) shows the powder X-ray diffraction (PXRD) patterns of pure and 0.7% Al-doped CrAs crystals. Rietveld refinement of the pattern matches the peak positions and intensities of the orthorhombic MnP-type crystal structure (space group *Pnma*, 62), indicating that the single crystals are in a single-phase without any detectable impurity phases. The lattice constants change from $a$=5.6510, $b$=3.4688 and $c$=6.2067 Å for pure CrAs to 5.6499, 3.4831 and 6.2049 Å for 0.7% Al-doped CrAs, showing that the $b$-axis lattice constant is elongated and cell volume increased by about 0.4% with Al doping. Here, the Al content in CrAs was determined by energy-dispersive X-ray spectroscopy (EDS).

The $a$-axis electrical resistivity $\rho$ of Al-doped CrAs is compared to that of pure CrAs in Fig. 2(c). With decreasing temperature, there is a sharp drop with hysteresis at 260 and 270 K ($=T_N$) for pure and Al-doped CrAs, respectively, which arises from the coincidence of AFM and isostructural volume expansion transitions [19, 20]. Aluminum substitution not only increases $T_N$ by 10 K, but also increases $\rho$ at 290 K from 169 to 197 μΩ·cm, due in part to the increase in disorder and higher transition temperature. Likewise, the residual resistivity $\rho_0$, estimated by extrapolating $\rho(T)$ from base temperature to 0 K, increases from 1.4 to 6.0 μΩ·cm and the residual resistivity ratio (RRR) decreases from 120 to 33, again signifying that disorder from Al substitution contributes significantly to the electron scattering. Concomitant with the resistivity results, as shown in Fig. 2(d), magnetic susceptibility measurements find that $T_N$ increases from 260 K for pure CrAs to 270 K for Al-CrAs, demonstrating that Al substitution is of bulk nature.

Though the larger cell volume of Al-CrAs is consistent with its higher magnetic/structural transition temperature, Al- doping is not a simple negative chemical pressure effect. Figure 3(a) and (b) comparatively show the pressure dependence of the magnetic and superconducting phase transition temperatures of Al-doped and pure CrAs, respectively. $T_N$ of Al-CrAs decreases gradually with initial pressure but drops rapidly for pressures higher than 4 kbar, similar to what happens in pure CrAs near 7 kbar. If Al were acting solely as a negative chemical pressure, the crossover to a steep decrease in $T_N$ should occur at a higher pressure in Al-CrAs. Nevertheless, a smooth extrapolation of $T_N(P)$ indicates that the magnetic transition reaches $T= 0$ near 4.5 kbar (= $P_C$), which is nearly half the critical pressure of 8 kbar for CrAs, even though its $T_N$ is 10 K lower.

The increase in $T_N$ from 260 to 270 K with Al-doping might reasonably be expected from a negative pressure effect because of the larger cell volume and particularly expanded $b$-axis. Since the Cr-3d states are more localized due to reduction in p-d mixing between Cr 3d and the anion p states, the already sizeable ordered moment (1.73 $\mu_B$) in CrAs should increase with Al doping as should $T_N$ [24]. The substantially lower $P_C$, however, indicates that the effect of Al-doping cannot be explained simply by a reduction in p-d mixing. As with itinerant antiferromagnetism in V-doped Cr, pressure and chemical doping play very different roles due both to impurity scattering and to changes in the electronic structure [25, 26]. This is likely as well to be the case in Al-CrAs, where Al introduces both disorder and additional carriers as V does in Cr. Interestingly, a few atomic percent V in Cr also substantially decreases the critical pressure of antiferromagnetic order and induces a very rapid drop in $T_N(P)$ as the critical pressure is approached [26]. Clearly, experimental and theoretical studies of the pressure-dependent electronic structure will be important to understand the microscopic role of Al-doping in CrAs.

Color contour plots of the temperature and pressure variation in $\rho$ of Al-doped and pure CrAs are respectively depicted in Fig. 3(a) and (b) on a semi-logarithmic scale, where the raw data are plotted in Fig. S1 in the Supplementary Information (SI). Magnetic/structural transition temperatures overlaid in the contour map are determined from a peak in $d\rho/dT$ (see Fig. S2 in the SI). The low-$T$ resistivity contour of CrAs, illustrated in Fig. 3(b), does not show an anomalous behavior except for a slight change in the slope of the resistivity contour near 8 kbar. In contrast, the large scattering region marked by red in Al-doped CrAs, as shown in Fig. 3(a), forms a narrow funnel that emerges from the projected $T = 0$ K critical point at 4.5 kbar ($=P_C$). When the system is away from the critical pressure, the electron scattering rate at low temperatures becomes smaller, indicating that the anomalously large scattering rate at $P_C$ arises from the critical magnetic fluctuations associated with the quantum phase transition.

Figure 4(a) gives the pressure evolution of the low-temperature resistivity of Al-CrAs, where $\rho$ is plotted as a function of $T$ and shifted rigidly with an offset at each pressure for clarity. The dashed red lines are least-squares fits to a power-law behavior, i.e., $\rho = \rho_0 + AT^n$, where the best results of residual resistivity $\rho_0$ and coefficient $A$ are plotted on the left and right ordinates of Fig. 4(b), respectively. At lower pressures ($P < P_C$), $n$ is close to 2, as expected for Landau-Fermi liquid behavior. With increasing pressure, the exponent $n$ of Al-CrAs sharply drops close to 1.5 at 4.5 kbar and gradually increases to 1.83 at 24.1 kbar. The non-Fermi liquid behavior can be ascribed to scattering by critical fluctuations associated with the AFM QCP at $P_C$. Underpinning the presence of the QCP, $\rho_0$ shows a sharp peak and $A$ abruptly increases by a factor of 30 at the critical pressure $P_C$ (=4.5 kbar), as shown in Fig. 4(b) and (c). In pure CrAs, analysis of the low-$T$ resistivity by power-law fits also shows that the non-Fermi liquid behavior appears near 8 kbar, the pressure across which there occurs a change in the slope of resistivity, and the critical region extends over a wide pressure range (P > 8 kbar) - see Fig. S3 in the SI. The fact that non-Fermi liquid behavior is observed near the tuned QCP of Al-CrAs underscores that the strange metallic behavior near the optimal pressure ($P_c$) in CrAs is originated from the critical magnetic fluctuations associated with the AFM QCP veiled by the dome of SC phase.

The SC transition temperature $T_c$, defined by the point of zero resistance in Al-doped CrAs, starts to appear for pressures above the critical pressure $P_C$ (=4.5 kbar), as marked by the triangles in Fig. 3(a). With further increasing pressure, $T_c$ reaches a maximum near 10 kbar, and gradually decreases to 1.27 K at 24.1 kbar. When compared with pure CrAs (see Fig. 3(b)), where the zero-resistance state starts to appear at 3 kbar, the pressure required to induce an initial zero-resistance state is shifted to a higher pressure in Al-CrAs. However, the pressure-dependent dome of $T_c(P)$ is similar to that of pure CrAs, and a broad maximum appears near 10 kbar for both compounds. We note that the zero-resistance state starts to appear deep in the AFM state for pure CrAs, while it exists only after the AFM phase is completely suppressed for Al-CrAs.

The superconductivity detached from a quantum critical point in Al-CrAs is in stark contrast with those unconventional superconductors where non-Fermi liquid behaviors appear above the optimal doping or pressure at which the highest $T_c$ appears. Superconductivity in pure CrAs has been proposed to be mediated by magnetic fluctuations because of its proximity to a possible AFM QCP [19-21]. The absence of a coherence peak in the spin-lattice relaxation rate $T_1$ and a $T^3$ dependence of $1/T_1$ below $T_c$ are consistent with unconventional superconductivity in CrAs [21]. The observations of both six-fold and two-fold symmetric components in the field-angle dependent upper critical field of CrAs support this conclusion and further argue for odd-parity spin-triplet pairing [27]. When CrAs is doped with Al, however, the SC dome as well as the maximum $T_c$ at the optimal pressure near 10 kbar are independent of the disorder even though the residual resistivity increases from 1.4 to 6.0 $\mu\Omega\cdot$cm with Al doping. This robustness of the superconductivity against introduction of non-magnetic impurities seems at odds with simple triplet superconductivity that is easily destroyed by any type of impurities [28]. Further, muon-spin rotation measurements of CrAs under pressure find scaling of the superfluid density, $n_s \propto T_c^{3.2}$, which is consistent with conventional phonon pairing [23]. The peculiar band structure of CrAs, where its possible non-trivial band crossing is protected by the non-symmorphic crystal structure, may be important in unraveling the mechanism and nature of superconductivity in this fascinating material [23, 29]. Additional study that can give direct information on the SC gap, such as point contact spectroscopy and field-directional specific heat measurements under pressure, will be important to resolve these contradicting results.

Figure 5(a) and (b) sequentially describes a color contour map of normalized isothermal resistivity in the $T$-$P$ plane for pure and Al-doped CrAs, where electrical resistivity $\rho(P)$ was divided by $\rho$ at the highest measuring pressure for comparison, e.g., n$\rho$ = $\rho(P)$ / $\rho$(24.1 kbar) for Al-CrAs. The representative raw data that Fig. 5 is based on are plotted in Fig. S4 in the SI. In the high-$T$ regime, strong electron scattering (red) is observed mainly outside the magnetic phase boundary due to thermally induced critical fluctuations for both compounds. In the low-$T$ regime of CrAs, as shown in Fig. 5(b), the area of enhanced scattering extends over the high-pressure regime ($P>P_c$), which is consistent with recent neutron scattering and NQR results that showed abundant magnetic fluctuations in the normal state near the $T_c$ maximum pressure [21, 22]. There occurs a rapid change in the resistivity across the critical pressure of 8 kbar. The suppression of $T_c$ at this critical pressure is indicative of a competition between SC and the double helical magnetic phase. On the other hand, the low-$T$ isothermal resistivity of Al-CrAs in Fig. 5(a) shows a disparate behavior, forming a funnel of strongly enhanced electron scattering in the vicinity of the critical pressure of 4.5 kbar. The enhanced scattering decreases gradually with increasing the distance below and above $P_C$ (see Fig. S4(a) in the SI). The fact that the transition width becomes broader at lower temperatures is in contrast to a classical phase transition where the phase transition becomes sharper with decreasing temperature due to reduction in thermal fluctuations. The lambda-like enhanced resistivity across the critical pressure indicates that nature of the magnetic transition is weakly $1^{st}$ order or $2^{nd}$ order at lower temperatures.

The pressure evolution of the double helical magnetic phase of CrAs, which is important to shed light on the nature of superconductivity, has yet to be elucidated [16, 17, 21, 22]. Shen et al. performed neutron diffraction on a polycrystalline sample and reported a spin reorientation transition from the *ab* plane to the *ac* plane at 6 kbar, whereas the pressure-induced magnetic phase disappears near 9.4 kbar [17]. In contrast, Matsuda et al. studied neutron diffraction on a single crystal CrAs under hydrostatic pressure and reported that the helical magnetic order completely disappears at 6.9 kbar [22]. This discrepancy on the spin reorientation transition was then ascribed to the usage of single crystals in their study. As shown in Fig. 5(b), enhanced resistivity extends across the critical pressure of 8 kbar, but it is difficult to find any clear signature that may indicate the presence of an additional phase transition near 9.4 kbar – see also Fig. S4(b) in the SI. Unlike pure CrAs, the enhanced isothermal resistivity in Al-CrAs is not extended above the critical pressure of 4.7 kbar, instead it is mostly confined near $P_c$ in Fig. 5(a). Absence of any clear signature at higher pressure indicates that it is unlikely to have an additional phase transition in Al-CrAs. In order to make a definitive statement on the spin reorientation transition in Al-CrAs, however, further work such as neutron scattering or NQR experiments under pressure is necessary.

## Summary

Observation of a quantum critical point veiled by a dome of SC phase has been key in interpreting the origin of non-Fermi liquid behaviors in the normal state, coexistence of competing phases, and unconventional superconductivity for various classes of correlated superconductors. By synthesizing 0.7% Al-doped CrAs single crystals whose AFM phase transition and isostructural volume expansion occur at 270 K, a 10 K increase from 260 K in pure CrAs, we successfully moved a critical point from 8 to 4.5 kbar. Observation of a funnel of enhanced electron scattering and the non-Fermi liquid behavior at the critical pressure is consistent with a magnetic quantum critical point at that pressure for Al-doped CrAs. The SC phase, in contrast, is almost independent of the Al doping and forms a dome as a function of pressure with a maximum $T_c$ near 10 kbar. The superconductivity detached from the quantum critical point in Al-doped CrAs is different from pure CrAs and suggests that Cooper pair formation is not mediated solely, or if at all, by critical magnetic fluctuations. These discoveries not only point to new directions and needs for future theory and experiment work on CrAs but also indicate more broadly the power of using multiple non-thermal tuning parameters simultaneously to reveal the relationship between superconductivity and a hidden quantum critical point in classes of unconventional superconductors.

**Material and methods**

Single crystals of pure CrAs and CrAs doped with 0.7% Al were grown out of a Sn-flux, as described elsewhere [30]. Powder X-ray diffraction was measured with a Rigaku miniflex-600 (Cu $K$-α source, λ ~ 1.5406 Å) and the data were refined in the Fullprof program to determine the lattice constants at room temperature. A conventional four-probe technique was applied to measure electrical resistivity of needle-shaped CrAs with current flow in the needle along its elongated crystalline $a$-axis. At ambient pressure, the residual resistivity ratio (RRR) of Al-CrAs is approximately 33, which is lower than that of pure CrAs. Pressure measurements to 24.1 kbar were performed in a hybrid Be-Cu/NiCrAl clamp-type pressure cell with silicone oil as the pressure medium. The pressure-dependent superconducting transition temperature of a Pb manometer was used to determine the pressure [31]. Resistivity under pressure was measured in CCR (closed cycle refrigerator) and $^3$He refrigerators for relatively high- (2.8~305 K) and low-temperature ranges (0.25~4 K), respectively. Magnetic susceptibility measurements were carried out in a Magnetic Property Measurement System (MPMS, Quantum Design, Inc.) in an applied field of 5 kOe.

**Acknowledgments**

This work was supported by a National Research Foundation (NRF) of Korea grant funded by the Korean Ministry of Science, ICT, and Planning (No. 2012R1A3A2048816). Work at Los Alamos National Laboratory was performed under the auspices of the U.S. Department of Energy, Office of Basic Energy Sciences, Division of Materials Sciences and Engineering.

**References**

[1] N. D. Mathur, F. M. Grosche, S. R. Julian, I. R. Walker, D. M. Freye, R. K. W. Haselwimmer and G. G. Lonzarich (1998) Magnetically mediated superconductivity in heavy fermion compounds. *Nature* 394:39-43.

[2] G. R. Stewart (2001) Non-Fermi-liquid behavior in d-and f-electron metals. *Rev Mod Phys* 73:797-855.

[3] H. Q. Yuan, F. M. Grosche, M. Deppe, C. Geibel, G. Sparn, F. Steglich (2003) Superconducting Phases in $CeCu_2Si_2$. *Science* 302:2104 -2107.

[4] T. Park, et al. (2006) Hidden magnetism and quantum criticality in the heavy fermion superconductor $CeRhIn_5$. *Nature* 440:65-68.

[5] G. Knebel, D. Aoki, D. Braithwaite, B. Salce, and J. Flouquet (2006) Coexistence of antiferromagnetism and superconductivity in CeRhIn5 under high pressure and magnetic field. *Phys Rev B* 74:020501-1.

[6] H. v. Lohneysen, A. Rosch and M. Vojta (2007) Fermi-liquid instabilities at magnetic quantum phase transitions. *Rev Mod Phys* 79:1015-1075.

[7] P. Monthoux, D. Pines and G. G. Lonzarich (2007) Superconductivity without phonons. *Nature* 450:1177-1183.

[8] P. Gegenwart, Q. Si and F. Steglich (2008) Quantum criticality in heavy-fermion metals. *Nature Phys* 4:186-197.

[9] T. Park, el al. (2008) Isotropic quantum scattering and unconventional superconductivity. *Nature* 456:366-368.

[10] D. J. Scalapino (2012) A common thread: The pairing interaction for unconventional superconductors. *Rev Mod Phys* 84:1383-1417.

[11] Y. Nakai, et al. (2010) Unconventional Superconductivity and Antiferromagnetism Quantum Critical Behavior in the Isovalent-Doped $BaFe_2(As_{1-x}P_x)_2$ *Phys Rev Lett* 105:107003.

[12] James G. Analytis, et al. (2014) Transport near a quantum critical point in $BaFe_2(As_{1-x}P_x)_2$ *Nat Phys* 10:194-197.

[13] B. Keimer, S. A. Kivelson, M. R. Norman, S. Uchida and J. Zaanen (2015) From quantum matter to high-temperature superconductivity in copper oxides *Nature* 518:179-186.

[14] K. Jin, N. P. Butch, K. Kirshenbaum, J. Paglione and R. L. Greene (2011) *Nature* 476:73-75.

[15] S. Seo, et al. (2015) Controlling superconductivity by tunable quantum critical point. *Nat Commun* 6:6433.

[16] L. Keller, et al. Pressure dependence of the magnetic order in CrAs:A neutron diffraction investigation. *Phys Rev B* 91:020409.

[17] Yao Shen, et al. (2016) Structural and magnetic phase diagram of CrAs and its relationship with pressure-induced superconductivity. *Phys Rev B* 93:060503.

[18] Z. Yu, et al. (2015) Anomalous anisotropic compression behavior of superconducting CrAs under high pressure. *Proc Natl Acad Sci USA* 112:14766-14770.

[19] W. Wu, et al. (2014) Superconductivity in the vicinity of antiferromagnetic order in CrAs. *Nat Commun* 5:5508.

[20] H. Kotegawa, S. Nakahara, H. Tou, and H. Sugawara (2014) Superconductivity of 2.2K under pressure in Helimagnet CrAs. *J Phys Soc Jpn* 83:093702.

[21] H. Kotegawa, et al. (2015) Detection of an Unconventional Superconducting Phase in the Vicinity of the Strong First-Order Magnetic Transition in CrAs Using $^{75}$As-Nuclear Quadrupole Resonance. *Phys Rev Lett* 114:117002.

[22] M. Matsuda, et al. (2018) Evolution of Magnetic Double Helix and Quantum Criticality near a Dome of Superconductivity in CrAs. *Phys Rev X* 8:031017.

[23] R. Khasanov, et al. (2015) Pressure-induced electronic phase separation of magnetism and superconductivity in CrAs. *Sci Rep* 5:13788.

[24] K. Motizuki, H. Ido, T. Itoh and M. Mortifuji (2009) Electronic structure and magnetism of 3d-transition metal pnictides. *Springer Ser Mater Sci* 131:1-142.

[25] T. M. Rice, JR. A. S. Barker, B. I. Halperin and D. B. McWhan (1969) Antiferromagnetism in Chromium and its Alloys. *J App Phys* 9:1337-1343.

[26] R. Jaramillo, et al. (2009) Breakdown of the Bardeen-Cooper-Schrieffer ground state at a quantum phase transition. *Nature* 459:405-409.

[27] C. Y. Guo, et al. (2018) Evidence for triplet superconductivity near an antiferromagnetic instability in CrAs. *Phys Rev B* 98:024520.

[28] A. P. Mackenzie, et al. (1998) Extremely Strong Dependence of Superconductivity on Disorder in $Sr_2RuO_4$. *Phys Rev Lett* 80:161-164.

[29] A. P. Schnyder and P. M. R. Brydon (2015) Topological surface states in nodal superconductors. J Phys: Condens Matter 27:243201.

[30] W. Wu, et al. (2010) Low temperature properties of pnictide CrAs single crystal. *Science Chania* 53**:**1207-1211.

[31] A. Eiling and J. S. Schilling (1981) Pressure and temperature dependence of electrical resistivity of Pb and Sn from 1-300 K and 0-10GPa–use as continuous resistive pressure monitor accurate over wide temperature range; superconductivity under pressure in Pb, Sn, and In. J. Phys F: Metal Phys 11:623-639.

**FIGURE LEGENDS**

**Fig. 1.** Schematic temperature-control parameter ($T$-$\delta$) phase diagram of superconductivity and quantum critical matter. (a) Unconventional superconductivity is often observed in proximity to a projected quantum critical point ($\delta_c$), where non-Fermi liquid behavior is observed in the normal state above the SC phase. (b) When $\delta_c$ is tuned to a new point $\delta_c'$, the SC phase can be either moved to the new $\delta_c'$ or remain at the old $\delta_c$ – see the main text for discussion.

**Fig. 2.** Powder X-ray diffraction and physical properties at ambient pressure. (a) Powder X-ray diffraction patterns of pure and 0.7% Al-doped CrAs in upper and lower panels, respectively. Inset to lower panel: magnified view near (102) and (111) peaks for both pure and Al-doped CrAs. (b) Schematic crystal structure of CrAs, where As and Cr atoms are depicted as green and blue spheres. (c) Electrical resistivity $\rho$ of pure (black symbols) and Al-doped CrAs (red symbols) for electrical current applied along the crystalline $a$-axis. The inset is an expanded view of $\rho$ near the transition temperature. Solid (open) symbols are data taken with decreasing (increasing) temperature. (d) Magnetic susceptibility ($\chi$) of pure (black symbols) and Al-doped CrAs (red symbols). The inset plots $\chi(T)$ near the AFM transition. Solid (open) symbols are data taken with decreasing (increasing) temperature.

**Fig. 3.** Three-dimensional contour map of electrical resistivity of Al-doped CrAs in (a) and pure CrAs in (b). The magnitude of the electrical resistivity is described by false colors in the temperature-pressure plane. (a) Anomalously enhanced electron scattering (red color) forms a funnel shape near the projected critical pressure of 4.5 kbar for Al-CrAs. Sold squares represent $T_N$, while solid triangles describe the zero-resistance SC transition temperature $T_c$. (b) $T_N$ of CrAs is described by solid and open squares that are data taken from this work and Ref. [19], respectively. Solid triangles represent $T_c$ of CrAs that are taken from Ref. [19]. $T_c$ of CrAs and Al-CrAs are multiplied by 1.3 for comparison. Dotted lines connected with AFM transitions are guides to the eyes. AFM and SC stands for antiferromagnetic and superconducting phase, respectively.

**Fig. 4.** Low-temperature electrical resistivity of Al-doped CrAs under pressure. (a) Electrical resistivity ($\rho$) of Al-doped CrAs is plotted against $T$ for several pressures. For clarity, $\rho$ at each pressure is shifted rigidly by an offset. Solid lines are least-squares fits to a power-law form, i.e. $\rho = \rho_0 + AT^n$, where the fitted values for $\rho_0$ (squares) and $A$ (circles) are plotted as a function of pressure on the left and right ordinates in (b), respectively. (c) The exponent $n$ (solid squares) is plotted on the left ordinate and the first derivative of the coefficient $A$ with respect to pressure ($P$), $dA/dP$ (solid circles), is shown on the right ordinate, where it is peaked near 4.5 kbar. Error

bars describe the standard deviation from the least-squares fitting.

**Fig. 5.** Three-dimensional contour map of normalized isothermal electrical resistivity (n$\rho$) as a function of pressure for Al-CrAs and CrAs in (a) and (b), respectively. For comparison, isothermal electrical resistivity of Al-CrAs is divided by $\rho$ at 24.1 kbar, while that of CrAs is divided by $\rho$ at 22.5 kbar, the highest measuring pressures for both compounds. Raw data on which the contour plot is based are plotted in Fig. S4 in the SI. Squares and triangles represent $T_N$ and $T_c$, respectively, where $T_c$ is multiplied by 20 for comparison. AFM and SC stands for antiferromagnetic and superconducting phase, respectively.

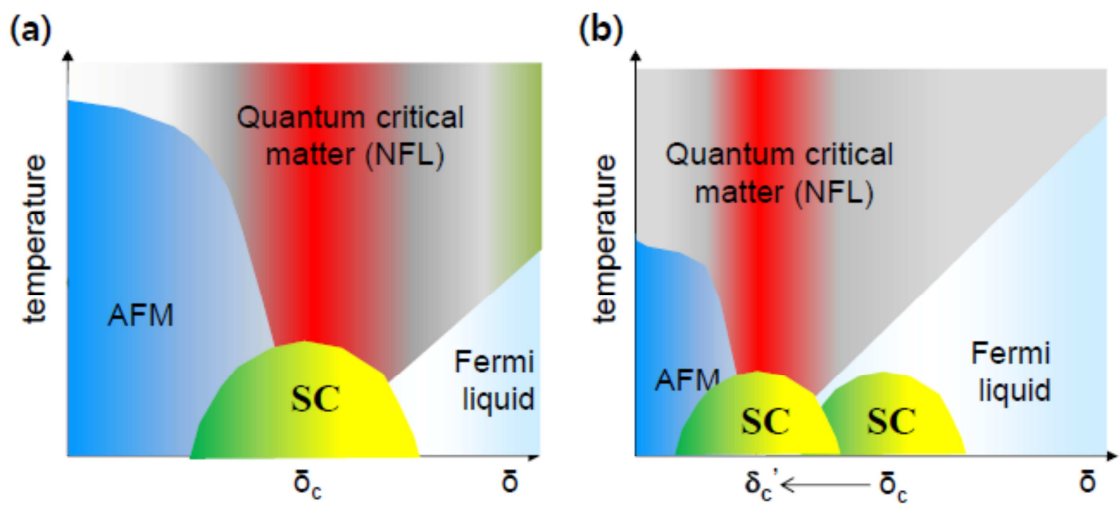

**Figure 1.**

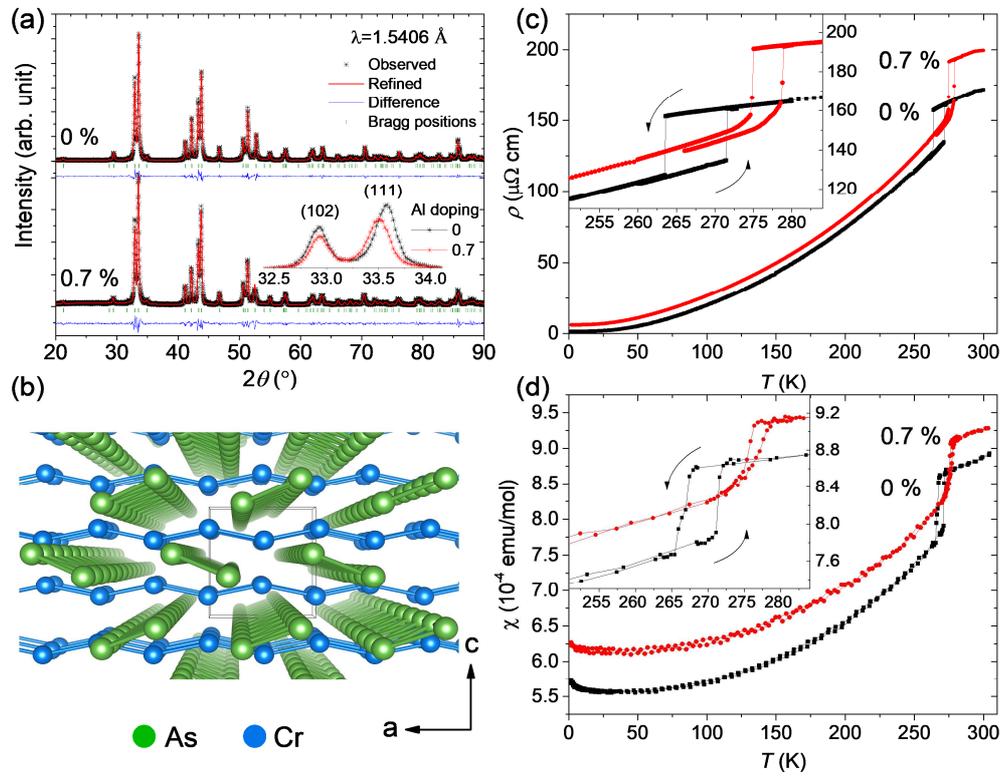

**Figure 2.**

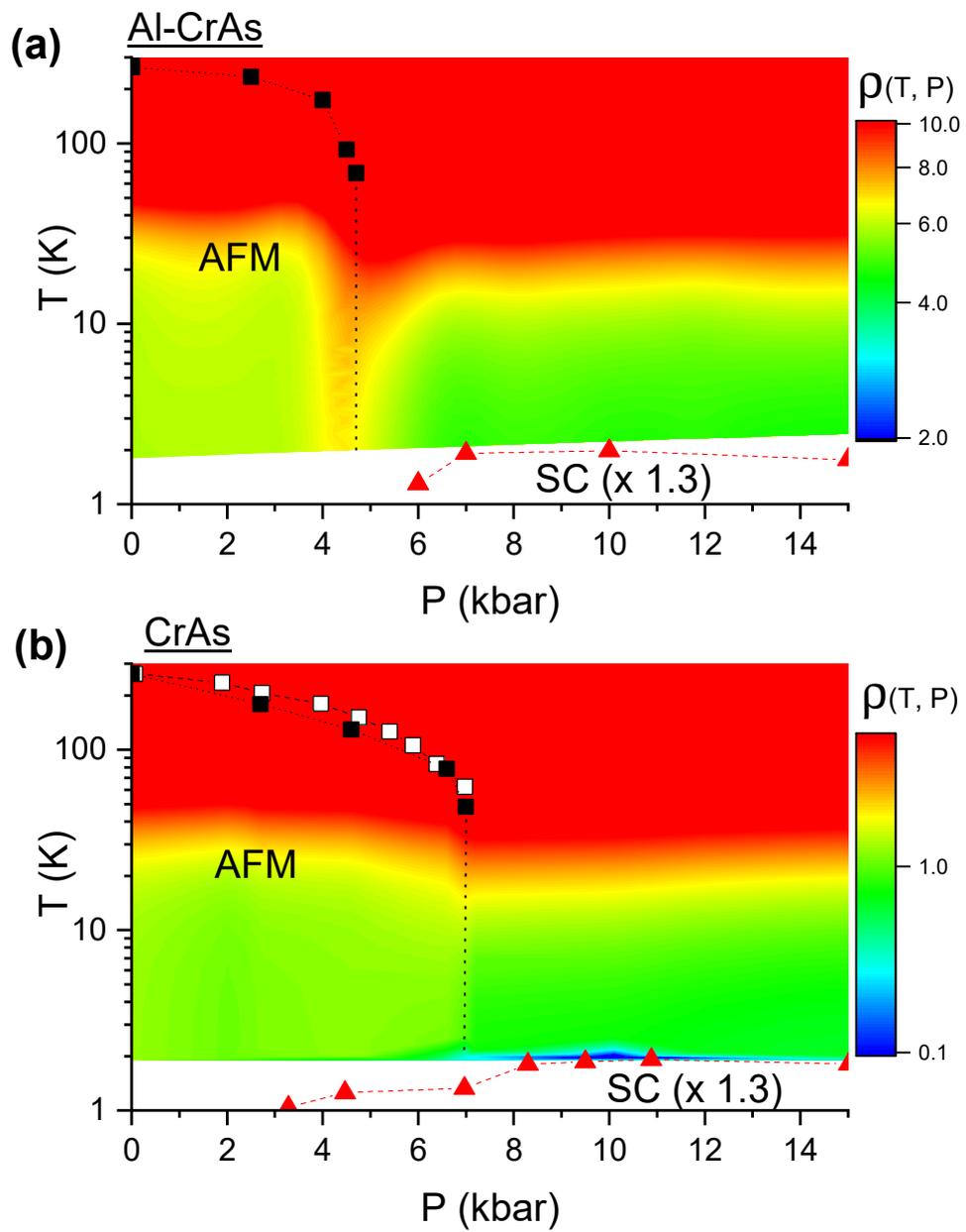

**Figure 3.**

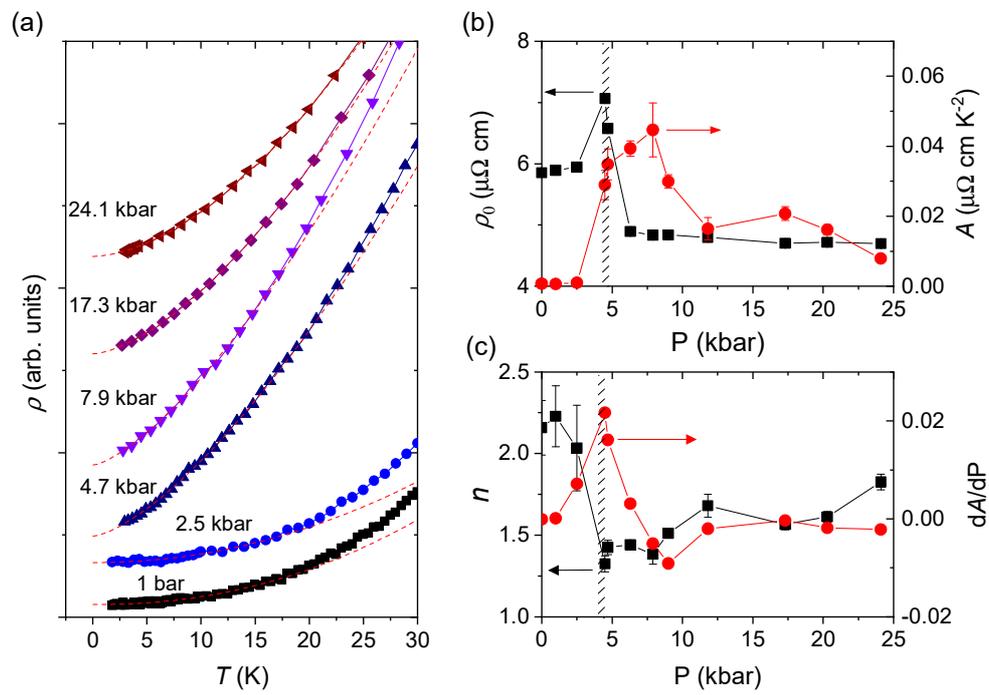

**Figure 4.**

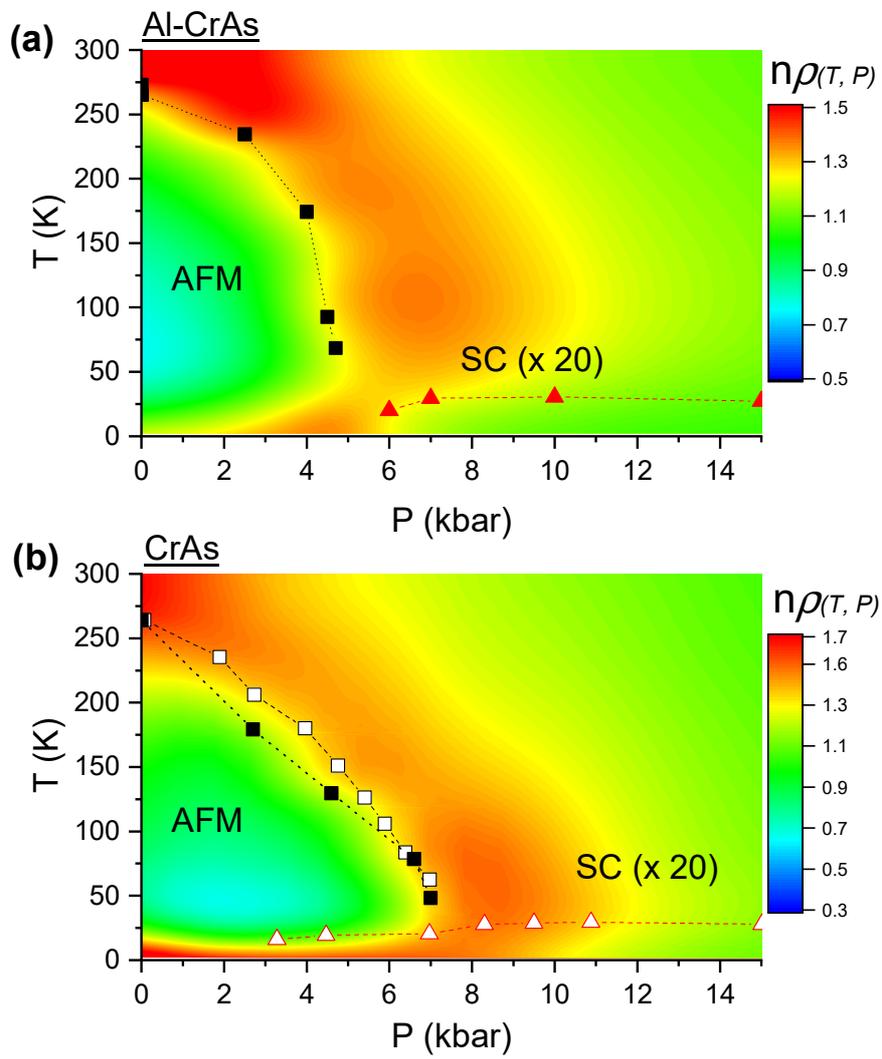

**Figure 5.**

# Supplementary information for Tunable quantum critical point and detached superconductivity in Al-doped CrAs

Sungmin Park[a,1], Soohyeon Shin[a,1], Sung-Il Kim[a], Joe D. Thompson[b], and Tuson Park[a,2]

[a]Department of Physics & Center for Quantum materials and Superconductivity Sungkyunkwan University, Suwon 16419, South Korea; [b]Los Alamos National Laboratory, Los Alamos, NM 87545, USA

[1]S. Park and S. Shin contributed equally to this work

[2]To whom correspondence should be addressed : Prof. Tuson Park, [a]Department of Physics & Center for Quantum materials and Superconductivity Sungkyunkwan University, Suwon 16419, South Korea

Tel. +82-31-299-4543, Email: tp8701@skku.edu

Classification:

PHSICAL SCIENCES: Physics

In this supplement, we present additional data that support results in the main text.

**Legends to supplementary figures**

**Fig. S1**. Electrical resistivity of Al-doped CrAs (Al-CrAs) and pure CrAs under pressure. Electrical resistivity of Al-CrAs and CrAs are plotted as a function of temperature in (a) and (b), respectively. Room-temperature resistivity gradually decreases with increasing pressure for both Al-CrAs and CrAs, which could be ascribed to the increased overlap between adjacent orbitals under pressure. A sharp drop in the resistivity at 1 bar occurs at 270 K (260

K) for Al-CrAs (CrAs) due to the AFM phase transition that is accompanied by isostructural volume expansion. With increasing pressure $T_N$ is suppressed for both compounds. Electrical resistivity data were taken with decreasing temperature.

**Fig. S2.** First temperature derivatives of electrical resistivity for several pressures. First derivative of electrical resistivity as a function of temperature for Al-CrAs and CrAs are shown in (a) and (b), respectively. A peak appears at the corresponding temperature of $T_N$, which is rapidly suppressed with increasing pressure.

**Fig. S3**. Power-law temperature dependence of the electrical resistivity of CrAs under pressure. (a) Electrical resistivity ($\rho$) of CrAs is plotted against $T$ for several pressures. For clarity, $\rho$ at each pressure is shifted rigidly by an offset. Solid lines are the least-squares fits to a power-law form, i.e. $\rho = \rho_0 + AT^n$, where the best results of $\rho_0$ (squares) and $A$ (circles) are plotted as a function of pressure on the left and right ordinates in (b), respectively. (c) The exponent $n$ (solid squares) is plotted on the left ordinate and the first derivative of the coefficient $A$ with respect to pressure ($P$), $dA/dP$ (solid circles), is shown on the right ordinate, where it is peaked near 8.0 kbar. Error bars describe the standard deviation from the least-squares fitting.

**Fig. S4.** Pressure dependence of the normalized isothermal electrical resistivity of Al-CrAs and CrAs. Normalized isothermal electrical resistivity is representatively plotted as a function of pressure for Al-CrAs and CrAs in (a) and (b), respectively. Here, the electrical resistivity of Al-CrAs is divided by that at 24.1 kbar, while the resistivity of CrAs is divided by that at 22.5 kbar. Arrows mark the critical pressure where resistivity shows a maximum due to the AFM phase transition for both compounds.

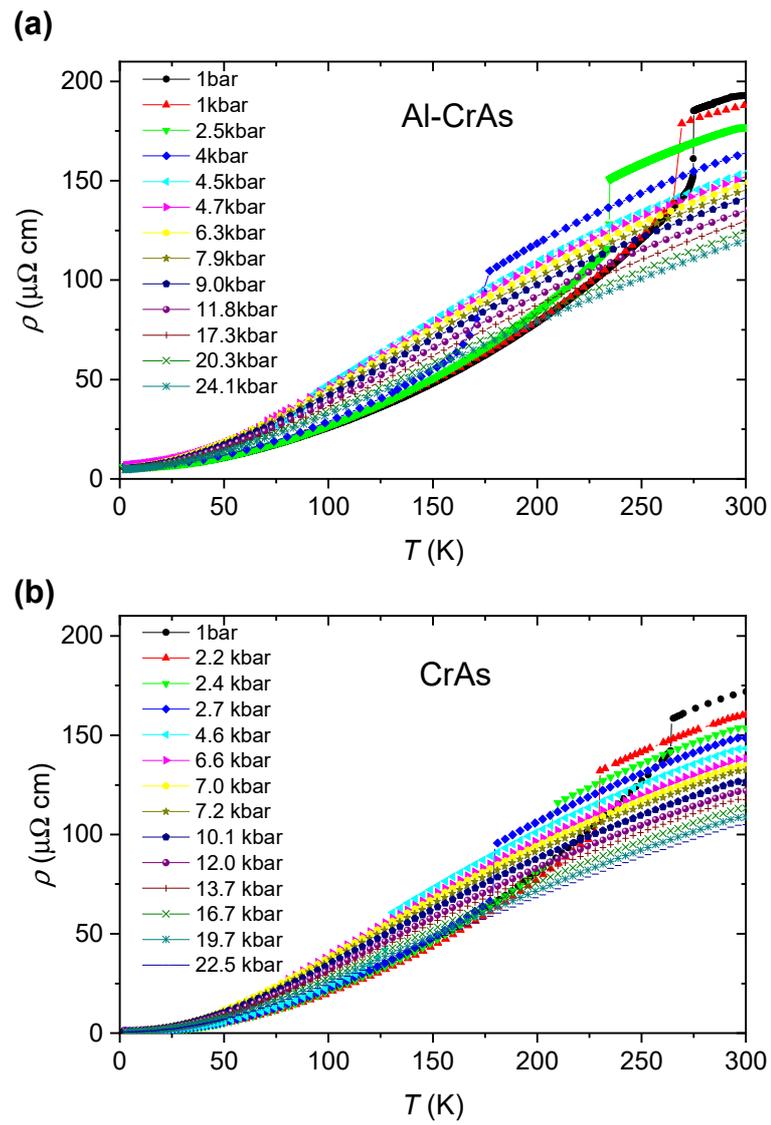

**Figure S1**.

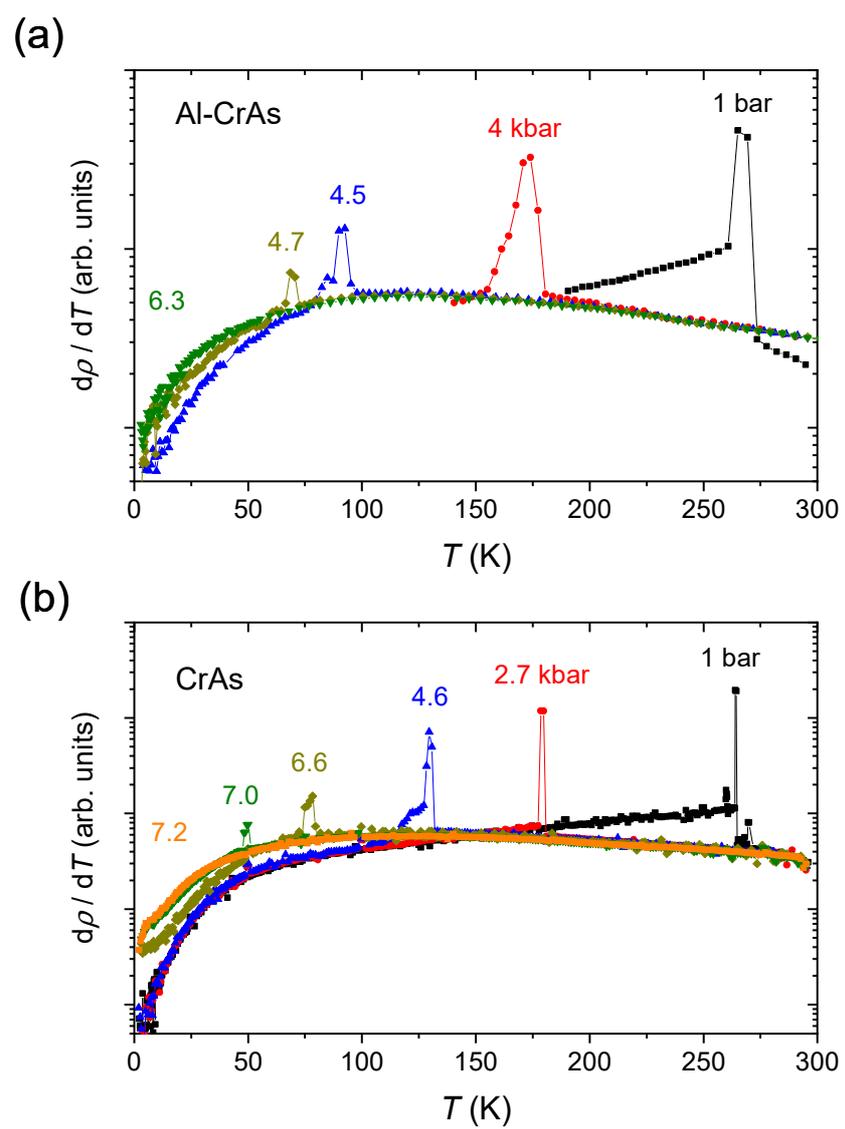

**Figure S2**.

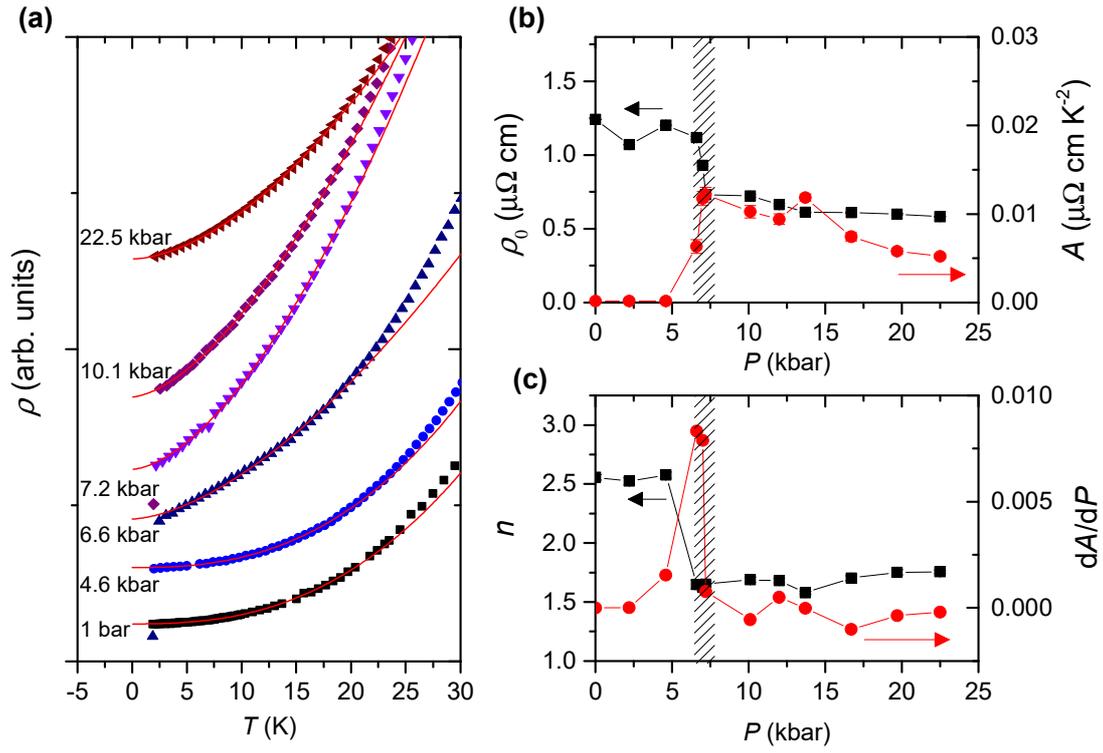

**Figure S3.**

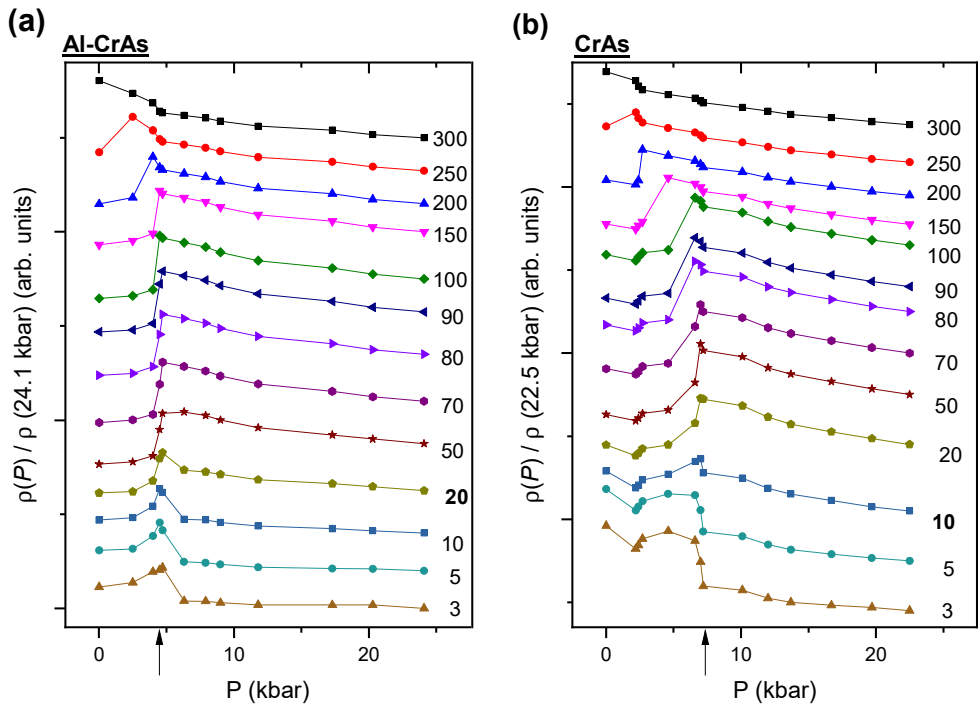

**Figure S4.**